\documentclass[singlecolumn,showpacs,preprintnumbers,amsmath,amssymb,floatfix]{revtex4}
\usepackage{graphicx}
\usepackage{dcolumn}
\usepackage{bm}

\begin{document}

\title{On the role of different Skyrme forces and surface corrections in exotic cluster-decay}

\author{Narinder K. Dhiman}%
 \email{narinder.dhiman@gmail.com}
\affiliation{%
Govt. Sr. Sec. School, Summer Hill, Shimla -171005, (India)\\
}%

\author{Ishwar Dutt}
\affiliation{
Department of Physics, Panjab University, Chandigarh -160014, (India)\\
}%


\date{\today}

\begin{abstract}

We present cluster decay studies of $^{56}$Ni$^*$ formed in
heavy-ion collisions using different Skyrme forces. Our study
reveals that different Skyrme forces do not alter the transfer
structure of fractional yields significantly. The cluster decay
half-lives of different clusters lies within $\pm$10\% for PCM and
$\pm$15\% for UFM.
\end{abstract}


\maketitle

\section{\label{intro}Introduction}
 Recently, a renewed interest has emerged in nuclear physics
research. This includes low energy fusion process~\cite{id},
intermediate energy phenomena~\cite{qmd} as well as cluster-decay
and/or formation of super heavy nuclei~\cite{gupta,kp}. In the
last one decade, several theoretical models have been employed in
the literature to estimate the half-life times of various exotic
cluster decays of radioactive nuclei. These outcome have also been
compared with experimental data. Most of these models applied to
study exotic cluster decay can be classified into two categories:
In the first category, only barrier penetration probabilities are
considered. Such models have been labeled as unified fission
models (UFM)~\cite{7Poen,7Buck,7Sand}. In the second category,
clusters are assumed to be formed well before penetration. This is
done by including the preformation probability in the
calculations. These models have been dubbed as preformed cluster
models (PCM)~\cite{rkg88,mal89,kum97}. In either of these
approach, one needs complete knowledge of the potential.
\par
This problem is tackled in the literature in two different
manners: One tries to adjust various parameters of model to known
experimental data~\cite{8MS,8CW,8Blocki77,8Bass}. Alternatively,
one starts from a basic fundamental approach free from such
adjustable
parameters~\cite{8Deni02,8Stancu,8Puri92,8Dob,8Brack,2Skyrme,VB72}.
It remain to be seen how particular set of model parameters
influence the cluster decay process. We plan to address this
question in this paper. We shall work out the above problem with
potential obtained from the Skyrme interactions. The Skyrme
interactions are well used to describe the fusion process at low
incident energies as well as in subthreshold, collective flow, and
multifragmentation at intermediate energies.
\par

The Skyrme force is an effective interaction, which parameterizes
the \emph{G-matrix} by a zero range, density and momentum
dependent ansatz. The Skyrme force consists of two-body as well as
three-body parts as~\cite{2Skyrme}:
\begin{equation}
V = \sum_{i< j} v_{ij} + \sum_{i< j< k} v_{ij k}. \label{eq:1}
\end{equation}
Using a short-range expansion of the two-body interaction, the
matrix elements in momentum space can be written as:
\begin{equation}
\langle \vec{k}\mid v_{12}\mid \vec{k}^{\prime} \rangle =
t_{0}\left(1 + x_{0}P_{\sigma} \right) +
\frac{1}{2}t_{1}\left(k^{2}+k^{\prime 2} \right)
+t_{2}\vec{k}\cdot \vec{k}^{\prime }+ i W_{0}\left(\vec{\sigma
}_{1} + \vec{\sigma}_{2}\right)\cdot \vec{k}\times
\vec{k}^{\prime},
 \label{eq:2}
\end{equation}
where $\vec{k}$ and $\vec{k}^{\prime}$ are the relative wave
vectors of the nucleons. $P_\sigma$ is spin exchange operator and
$\vec{\sigma}$ are Pauli spin matrices. To deal with such
interaction, it is convenient to write the matrix elements in
configuration space as:
\begin{eqnarray}
v_{12} &= &t_{0}\left(1+ x_{0}P_{\sigma }
\right)\delta\left(\vec{r}_{1}-\vec{r}_{2}\right)+
\frac{1}{2}t_{1}
\left[\delta\left(\vec{r}_{1}-\vec{r}_{2}\right)k^{2}+ k^{\prime
2}\delta\left(\vec{r}_{1}-\vec{r}_{2}\right)\right] \nonumber \\
& &+
t_{2}\vec{k}^{\prime}\cdot\delta\left(\vec{r}_{1}-\vec{r}_{2}\right)\vec{k}
+ i
W_{0}\left(\vec{\sigma}_{1}+\vec{\sigma}_{2}\right)\cdot\vec{k}^{\prime}\times
\delta\left(\vec{r}_{1}-\vec{r}_{2}\right)\vec{k},
 \label{eq:3}
\end{eqnarray}
here $ \vec{k}(=(\vec{\nabla }_{1} - \vec{\nabla }_{2} )/2\iota)$
denotes the relative momentum operators acting on the right and $
\vec{k}^\prime (= - (\vec{\nabla }_{1} - \vec{\nabla }_{2}
)/2\iota)$, acting on left, respectively.
\par

The three-body term of the Skyrme force can be written as:
\begin{equation}
v_{123} = t_{3}\delta\left(\vec{r}_{1}-\vec{r}_{2}\right)
\delta\left(\vec{r}_{2}-\vec{r}_{3}\right). \label{eq:4}
\end{equation}
For the Hartree-Fock calculations of even-even nuclei, this force
is shown to be equivalent to a two-body density dependent
interaction:
\begin{equation}
v_{12} =\frac{1}{6}t_{3}\left(1+ P_{\sigma }
\right)\delta\left(\vec{r}_{1}-\vec{r}_{2}\right)\rho\left(\frac{\vec{r}_{1}+\vec{r}_{2}}{2}\right).
\label{eq:5}
\end{equation}
The above form,  Eq.~(\ref{eq:5}), provides a simple
phenomenological representation of many body effects describing
the way, in which the interaction between two nucleons is
influenced by the presence of others. The Skyrme interaction is an
approximate representation of the effective nucleon force which is
valid only for the low relative momentum. In Eqs.~(\ref{eq:2}) to
(\ref{eq:5}), we see several constants/parameters like $t_0$,
$t_1$, $t_2$, $t_3$, $x_0$, and $W_0$ that need to be fitted.
These parameters have been fitted by various authors from time to
time to get better description of various ground state properties
of nuclei~\cite{8Puri92,2Skyrme,VB72,shen09}. A particular set
comprising these parameters is known as \emph{Skyrme force}. Till
to-date, large number of Skyrme forces are available in the
literature~\cite{shen09}. These different Skyrme forces
constituting different equation of state at intermediate
energies\cite{sk,rk}.
 All the conventional (i.e. with the three
body term replaced by a density dependent two body term),
generalized (adjusting the effective mass $m^*$ and
compressibility $K$) and modified Skyrme forces (adjusting the
density parameter $t_3$ to fit the spectra) are unified in a
single form by Zhuo~\cite{Li91} as an extended Skyrme force.
\begin{equation}
V_{ES} = \sum_{i< j} v_{ij}. \label{eq:6}
\end{equation}
\par
Our aim here is to study the role of various Skyrme forces and
surface corrections in the exotic cluster decay process. This
study is still missing in the literature.
\par

In recent years, there have been a number of experimental and
theoretical
studies~\cite{Sanders99,Sanders88,Sanders91,Nouicer99,Beck01,Thumm01,Bhatt02,Sanders86,Betts,Sanders89,Sanders94,nkd03,MKS00}
aimed at understanding the decay of light compound nucleus formed
through heavy-ion reactions. In most of the reactions studied,
whereas the general conclusion about the formation probability for
the compound nucleus and characteristic features of its decay are
debated in terms of either fusion-fission
mechanism~\cite{Sanders99,Sanders91,mat97}, which may be
considered as  the emission of complex (or intermediate mass)
fragments, or a deep inelastic (DI) orbiting~\cite{shiva87}
mechanism  behaviour.
\par

One of such system is the doubly magic $^{56}$Ni, which is studied
by using several entrance channels ($^{16}$O + $^{40}$Ca, $^{32}$S
+ $^{24}$Mg, $^{28}$Si + $^{28}$Si) and at different incident
energies (1.5 to 2.2. times Coulomb
barrier)~\cite{Sanders99,Sanders88,Sanders91,Nouicer99,Beck01,Thumm01,Bhatt02,Sanders86,Betts,Sanders89,Sanders94}.
At these incident energies, the incident flux get trapped that
results in the formation of compound nucleus, which is in addition
to a significant large angle scattering cross-sections. For light
masses (A$<$44), the compound nucleus decays by the emission of
light particles  and $\gamma$-rays. An experimental measure of
this so called particle evaporation residue is the compound
nucleus fusion cross-section. For heavier systems, such as
$^{56}$Ni, a significant decay strength to heavier fragments is
also observed which could apparently not arise from a direct
reaction mechanism because of large mass asymmetry differences
between the entrance and exit channels. The measured angular
distributions and energy spectra are consistent with fission like
decays of the respective compound systems.
\par

The measured mass distribution for $^{56}$Ni shows a preferential
decays to channels comprising $\alpha$-nuclei $^{16}$O, $^{20}$Ne,
$^{24}$Mg and $^{28}$Si, and their complimentary
fragments~\cite{Betts,Sanders89,Sanders94}, independent of the
entrance channel nuclei and centre-of-mass energy $E_{cm}$. Such
an $\alpha$-structure is associated with the shell effects in the
potential energy surface of the compound nucleus~\cite{Sanders89},
though these are almost zero at the compound nucleus excitation
energies involved. Such an $\alpha$-nucleus structure in the
measured mass distribution of $^{56}$Ni has its origin in the
macroscopic energy~\cite{MKS00}.
\par

Cluster decay is studied for  $^{56}$Ni, when formed as an excited
compound system in heavy-ion collisions. Since $^{56}$Ni has
negative $Q_{out}$, and hence stable against both fission and
cluster decay processes. However, if is is produced in heavy-ion
reactions depending on the incident energy and angular momentum,
the excited compound system could either fission, decay via
cluster emissions or results in resonance phenomenon. The negative
$Q_{out}$ is different for various exit channels and hence would
decay only if it were produced with sufficient compound nucleus
excitation energy $E^{\ast}_{CN}~(=E_{cm} + Q_{in})$, to
compensate for negative $Q_{out}$, the deformation energy of the
fragments $E_d$, their total kinetic energy ($TKE$) and the total
excitation energy ($TXE$), in the exit channel as:
\begin{equation}
E^{\ast}_{CN} = \mid Q_{out} \mid + E_{d} + TKE + TXE.
 \label{eq:7}
\end{equation}
(see Fig.~1, where $E_d$ is neglected because the fragments are
considered to be spherical).  Here $Q_{in}$ adds to the entrance
channel kinetic energy $E_{cm}$ of the incoming nuclei in their
ground states.
\par

Section \ref{model} gives some details of the Skyrme energy
density model and preformed cluster model and its simplification
to unified fission model. Our calculations for the decay half-life
times of $^{56}$Ni$^*$ compound system and a discussion of the
results are presented in Section \ref{result}. Finally, the
results are summarized in Section \ref{summary}.

\section{\label{model} Model}
\subsection{Skyrme Energy Density Model}
In the Skyrme Energy Density Model (SEDM), the real part of
interaction potential $V_N(R)$ is defined as difference between
energy expectation value $E$ of the whole system calculated at a
finite distance $R$ and at infinity~\cite{8Puri92,VB72}.
\begin{equation}
V_{N}\left(r \right)  = E\left(r \right)- E\left(\infty  \right),
\label{eq:8}
\end{equation}
with
\begin{equation}
E = \int H \left(\vec{r} \right)\vec{dr}. \label{eq:9}
\end{equation}
In this formalism, the energy density functional $H \left(\vec{r}
\right)$ read as;
\begin{eqnarray}
H(\rho,\tau,\vec{J}) & = & \frac{\hbar^2}{2m} \tau +\frac{1}{2}t_0
[(1+
\frac{1}{2}x_0)\rho^2-(x_0+\frac{1}{2})(\rho_n^2+\rho_p^2)]+\frac{1}{4}(
t_1+t_2)\rho\tau \nonumber\\
& & +\frac{1}{8}(t_2-t_1)(\rho_n \tau_n+\rho_p \tau_p)
+\frac{1}{16} (t_2-3t_1) \rho \nabla^2 \rho \nonumber \\ & &
+\frac{1}{32} (3t_1+t_2) (\rho_n \nabla^2 \rho_n+\rho_p \nabla^2
\rho_p) + \frac{1}{4}t_3 \rho_n \rho_p \rho \nonumber \\ & &
-\frac {1}{2} W_0(\rho \vec{\nabla}\cdot\vec{J} +\rho_n
\vec{\nabla}\cdot\vec{J}_n+ \rho_p \vec{\nabla} \cdot\vec{J}_p) .
\label{eq:10}
\end{eqnarray}
Here $\rho=\rho_{n} + \rho_{p}$ is the nucleon density taken to be
two-parameter Fermi density and $\vec{J} = \vec{J}_{n} +
\vec{J}_{p}$ is the spin density which was generalized by Puri et
al.~\cite{8Puri92}, for spin-unsaturated nuclei. The remaining
term is the kinetic energy density $\tau=\tau_{n} + \tau_{p}$. The
Coulomb effects are neglected in the above energy density
functional, but will be added explicitly. In Eq.~(\ref{eq:10}),
six parameters $t_0$, $t_1$, $t_2$, $t_3$, $x_0$, and $W_0$ are
fitted by different authors to obtain the best description of the
various ground state properties for a large number of nuclei. As
discussed in the introduction, these different parameterizations
have been labeled as S, SI, SII, SIII etc..
\par
The evaluation of kinetic energy density term was done within the
Thomas-Fermi (TF) approximation which is a well known alternative
to the Hartree-Fock method. As shown by various
authors~\cite{gupta85}, the kinetic energy density $\tau$ can be
separated into volume term $\tau_{0}$ and surface term plus
reminder. In other words,
\begin{equation}
\tau = \tau_{0} + \tau_{\lambda}+...... \label{eq:11}
\end{equation}
In the first order approximation, one can limit to $\tau_{0}$ term
only. The volume term $\tau_{0}$ in this approximation is given by
\begin{equation}
\tau_{0} =  \frac{3}{5}
\left(\frac{3}{2}\pi^{2}\right)^{\frac{2}{3}}\rho^\frac{5}{3}.
\label{eq:12}
\end{equation}

The kinetic energy density $\tau$~\cite{gupta85,2Von}, after
including additional surface effects is
\begin{equation}
\tau =\tau_{0}+ \lambda \frac{\left(\vec{\nabla}
\rho\right)^2}{\rho},
 \label{eq:13}
\end{equation}
here, $\lambda$ is a constant  whose value has been a point of
controversy and different authors have suggested different values,
lying between $1/36$ and $9/36$. The above Thomas-Fermi
approximation for $\tau$ reduces the dependence of energy density
$H(\vec{r})$ to nucleon density $\rho$ only. The exchange effects
due to anti-symmetrization can be assimilated to reasonable extent
when Eq.~(\ref{eq:13}) is used~\cite{8Puri92}. We apply the
standard Fermi mass density distribution for nucleonic density:
\begin{equation}
\rho_i \left(R \right)=\frac{\rho_{0i} }{ 1+ \exp\left\{\frac{R -
R_{0i} }{a_i } \right\} },~~~~~~~~~~~~ - \infty \leq R \leq \infty
\label{eq:14}
\end{equation}
    The average central density $ \rho_{0i}$ given by \cite{8Stancu}
\begin{equation}
\rho_{0i} =\frac{3A_i}{4 \pi
R^{3}_{0i}}\frac{1}{\left[1+\frac{\pi^{2}a^{2}_i}{R^{2}_{0i}}
\right]}, \label{eq:15}
\end{equation}
$R_{0i}$  and a$_i$ are, respectively, the half-density radii and
surface diffuseness parameters taken from
Refs.\cite{8Puri92,Elton}. For the details of the model, reader is
referred to Ref. \cite{8Puri92}.
\par

\subsection{The Preformed Cluster Model}
For the cluster decay studies, we use the Preformed Cluster Model
(PCM)~\cite{rkg88,mal89,kum97}. This model, based on the quantum
mechanical fragmentation theory~\cite{7Puri,rkg75,7SNG,7Maruhn74},
uses the decoupled approximation to $\eta$- and $R$-motions. The
decay constant ($\Lambda$) in the PCM is defined as,
\begin{equation}
\Lambda = \nu _{0}PP_{0},\qquad\qquad\qquad\left(~{\rm
or}~~T_{1/2}=\frac{\ln 2}{\Lambda}~\right),
 \label{eq:16}
\end{equation}
here $\nu_0$ is the assault frequency with which the cluster hits
the barrier, $P$ is the probability of penetrating the barrier and
$P_0$ is the preformation probability. Thus, in contrast to the
unified fission models~\cite{7Poen,7Buck,7Sand}, the two fragments
in PCM are considered to be formed at a relative separation
co-ordinate $R$ before the penetration of the potential barrier
with probability $P_0$. The Schr\"odinger equation in terms of
$\eta$ and $R$ coordinates as:
\begin{equation}
H(\eta ,R)\psi (\eta ,R)=E\psi (\eta ,R),
 \label{eq:17}
\end{equation}
The above equation can be solved in a decoupled
approximation~\cite{rkg88,mal89}, for which the Hamiltonian takes
the form:
\begin{equation}
H = -\frac{\hbar ^{2}}{2\sqrt{B_{\eta\eta}}}\frac{\partial
}{\partial \eta }\frac{1}{\sqrt{B_{\eta \eta }}}\frac{\partial
}{\partial \eta } -\frac{\hbar^{2}}{2\sqrt{B_{RR }}}\frac{\partial
}{\partial R }\frac{1}{\sqrt{B_{RR }}}\frac{\partial }{\partial
R}+V(\eta)+V(R). \label{eq:18}
\end{equation}
Since the potentials are calculated within the Strutinsky
re-normalization procedure ($V = V_{Macro} + \delta U$) by using
an appropriate liquid drop model potential $V_{Macro}$ and
asymmetric two center shell model for shell corrections $\delta
U$, are nearly independent of the relative separation coordinate
$R$, $R$ can be taken as a time independent parameter. For the
Hamiltonian Eq.~(\ref{eq:18}), the Schr\"odinger Eq.~(\ref{eq:17})
can be separated in two co-ordinates $\eta$ and $R$ as follows:
\begin{equation}
\left[-\frac{\hbar ^{2}}{2\sqrt{B_{\eta\eta }}}\frac{\partial
}{\partial \eta }\frac{1}{\sqrt{B_{\eta \eta }}}\frac{\partial
}{\partial \eta }+V(\eta) \right]\psi(\eta)=E_\eta\psi(\eta),
\label{eq:19}
\end{equation}
and
\begin{equation}
\left[-\frac{\hbar ^{2}}{2\sqrt{B_{RR }}}\frac{\partial }{\partial
R }\frac{1}{\sqrt{B_{RR }}}\frac{\partial }{\partial R }+V(R)
\right]\psi(R)=E_R\psi(R),
 \label{eq:20}
\end{equation}
with $\psi(\eta, R)=\psi(\eta)\,\psi( R)$ and $E=E_{\eta} + E_R.$
\par
The fragmentation potential (or collective potential energy)
$V(\eta)$, appearing in Eq.~(\ref{eq:19}), is calculated as,
\begin{equation}
V(\eta)=-\sum^2_{i=1} \left[V_{Macro}(A_i,Z_i)+\delta U_i
\exp\left(-\frac{T^2}{T_0^2}\right)\right] +\frac{Z_1 \cdot
Z_2e^2}{R} + V_N(R) + V_{\ell},
 \label{eq:21}
\end{equation}
where the theoretical binding energies ($V = V_{Macro} + \delta
U$) are taken from M\"oller et al.~\cite{mol95}. The charges $Z_i$
in Eq.~(\ref{eq:21}) are fixed by minimizing the potential
$V(\eta_Z)$, defined by Eq.~(\ref{eq:21}) without $V_N(R)$ in
$\eta_Z$ co-ordinates. The shell corrections $\delta U$ are
considered to vanish exponentially for $E^{\ast}_{CN} \ge 60$ MeV,
giving $T = 1.5$ MeV. At higher excitation energies, the shell
corrections vanish completely and only the liquid drop part of
energy is present. The additional attraction due to nuclear
interaction potential $V_N(R)$ is calculated within SEDM
potential. The rotational energy due to angular momentum effects
$V_{\ell}~(= \hbar^2\ell(\ell+1)/2\mu R^2)$ is not added here
since its contribution to the structure yields is shown to be
small for lighter systems~\cite{7SNG}. The nuclear temperature $T$
(in MeV), is related approximately to the excitation energy
$E^{\ast}_{CN}$, as:
\begin{equation}
E^{\ast}_{CN}=\frac{1}{9}A{T}^2-T \qquad\qquad {(\rm in~ MeV)}.
\label{eq:22}
\end{equation}
\par

The kinetic energy part of the Hamiltonian in Eq.~(\ref{eq:19})
comes through the mass parameter $B_{\eta\eta}$ which is
calculated using the classical mass parameter of Kr\"oger and
Scheid~\cite{kro80}, based on the hydrodynamical flow. The mass
parameter $B_{\eta\eta}$ reads as:
\begin{equation}
B_{\eta \eta}=\frac{AmR^{2}_{min}}{4}\left[\frac{v_{t} (1+\beta
)}{v_{c}(1+\delta ^{2})-1} \right],
 \label{eq:23}
\end{equation}
with
\begin{equation}
\beta=\frac{R_{c}}{2R_{min}}\left[\frac{1}{1+\cos \theta
_{1}}\left(1-\frac{R_{c}}{R_{1}} \right)+ \frac{1}{1+\cos \theta
_{2}}\left(1-\frac{R_{c}}{R_{2}} \right) \right],
 \label{eq:24}
\end{equation}
\begin{equation}
\delta =\frac{1}{2R_{min}}\left[(1-\cos \theta_{1})(R_{1}-R_{c})
+(1-\cos \theta _{2})(R_{2}-R_{c}) \right],
 \label{eq:25}
\end{equation}
\begin{equation}
v_{c}= \pi R^{2}_{c}R_{min},~~~~~~~~~~~~~~~R_{c}= 0.4R_{2},
\label{eq:26}
\end{equation}
and $v_t=v_1 + v_2$, is the total conserved volume.
\par
Solving Eq.~(19) numerically, $\mid\psi(\eta)\mid^{2}$ gives the
probability of finding the mass fragmentation $\eta$ at a fixed
position $R$, on the decay path. Normalizing and scaling
$\mid\psi(\eta )\mid^{2}$ to give the fractional mass yield for
each fragment in the ground state decay as:
\begin{equation}
P_{0}(A_{i})= \mid \psi(\eta) \mid^{2}\sqrt{B_{\eta \eta }(\eta
)}\left(\frac{4}{A_i}\right),\,\,\,\,\,\,\,\,\,(i=1~{\rm or}~2).
 \label{eq:27}
\end{equation}
The nuclear temperature effects in Eq.~(\ref{eq:27}) are also
included through a Boltzmann-like function,
\begin{equation}
\mid \psi(\eta)\mid^{2}=\sum_{\nu=0}^{\infty }\mid \psi (\eta)
\mid^{2}\exp \left(-\frac{E_{\eta}}{T} \right). \label{eq:28}
\end{equation}
\par
For $R$-motion, instead of solving the stationary Schr\"odinger
Eq.~(\ref{eq:20}), the WKB action integral was solved for the
penetration probability $P$~\cite{7Rkg94}. For each $\eta$-value,
the potential $V(R)$ is calculated by using SEDM for $R \ge
R_{d}$, with $R_{d}=R_{min} + \Delta R$ and for $R \le R_{d}$, it
is parameterized simply as a polynomial of degree two in $R$:
\begin{equation}
V(R)=\left \{
\begin{array}{ll}
\mid Q_{out} \mid +{a_1}(R-R_0)+{a_2}(R-R_0)^2 &  \mbox{for \quad $R_0\leq R\leq R_{d} $}, \\
V_N(R) + Z_1 \cdot Z_2 e^2/R & \mbox{for $ \quad R\geq R_{d}$},
\end{array}
\right. \label{eq:29}
\end{equation}
where $R_0$ is the parent nucleus radius and $\Delta R$ is chosen
for smooth matching between the real potential and the
parameterized potential (with second-order polynomial in $R$). A
typical scattering potential, calculated by using Eq.~(29) is
shown in Fig.~1, with tunneling paths and the characteristic
quantities also marked. Here we choose the first (inner) turning
point $R_a$ at the minimum configuration i.e. $R_a = R_{min}$
(corresponding to $V_{min}$) with potential at this $R_a$-value as
$V(R_a = R_{min})= \overline{V}_{min}$ (displayed in Fig.~1) and
the outer turning point $R_b$ to give the $Q_{eff}$-value of the
reaction ($Q_{eff}= \mid Q_{out} \mid + TKE$) i.e. $V(R_b) =
Q_{eff}$. This means that the transmission probability $P$ with
the de-excitation probability, $W_i=\exp (-bE_i)$ taken as unity,
can be written as:
\begin{equation}
P=P_iP_b,
 \label{eq:30}
\end{equation}
where $P_i$ and $P_b$ are calculated by using WKB approximation,
as:
\begin{equation}
P_i=\exp \left[- \frac{2}{\hbar} \int\limits_{R_a}^{R_i}\{2\mu
[V(R)-V(R_i)]\}^{1/2}dR \right],
 \label{eq:31}
\end{equation}
and
\begin{equation}
P_b=\exp\left[- \frac{2}{\hbar} \int\limits_{R_i}^{R_b}\{2\mu
[V(R)-Q_{eff}]\}^{1/2}dR \right],
 \label{eq:32}
\end{equation}
here $R_a$ and $R_b$ are, respectively, the first and second
turning points. This means that the tunneling begins at $R =
R_a~(=R_{min})$ and terminates at $R = R_b$, with $V(R_b) =
Q_{eff}$. The integrals of Eqs.~(31) and~(32) are solved
analytically by parameterizing the above calculated potential
$V(R)$.
\par
The assault frequency or the barrier impinging frequency $\nu_0$
in Eq.~(16), is given simply as,
\begin{equation}
\nu_0 = \frac{v}{R_0} = \frac{(2E_2/\mu)^{1/2}}{R_0},
\label{eq:33}
\end{equation}
where $E_2 = \frac{A_1}{A} Q_{eff}$ is the kinetic energy of the
emitted cluster, with $Q_{eff}$ shared between the two fragments
and $\mu =m(\frac{A_1 A_2}{A})$ is the reduced mass.
\par

The PCM can be simplified to unified fission model (UFM), if
preformation probability $P_0=1$ and the penetration path is
straight to $Q_{eff}$-value.

\section{\label{result}Results and Discussions}
 The calculations are made in two steps: In the first steps, we studied
the role of different Skyrme forces in the cluster decay of
$^{56}$Ni$^{\ast}$ and  in the second step, effect of surface
correction term $\lambda$ is analyzed.
\par

Fig.~1 shows the characteristic scattering potential for the
cluster decay of $^{56}$Ni$^{\ast}$ into $^{16}$O + $^{40}$Ca
channel as an illustrative example.  In the exit channel for the
compound nucleus to decay, the compound nucleus excitation energy
$E_{CN}^{\ast}$ goes in compensating the negative $Q_{out}$, the
total excitation energy $TXE$ and total kinetic energy $TKE$ of
the two outgoing fragments as the effective Q-value (i.e.
$TKE=Q_{eff}$ in the cluster decay process). In addition, we plot
the penetration paths for PCM and UFM. For PCM, we begin the
penetration path at $R_a = R_{min}$ with potential at this
$R_a$-value as $V(R_a = R_{min})= \overline{V}_{min}$ and ends at
$R = R_b$, corresponding to $V(R=R_b) = Q_{eff}$, whereas for UFM,
we begin at $R_a$ and end at $R_b$ both corresponding to $V(R_a)
=V(R_b)=Q_{eff}$. We have chosen only the case of different
$Q_{eff}$ (listed in Table~1), for different cluster decay
products to satisfy the arbitrarily chosen relation
$Q_{eff}=0.4(28 - \mid Q_{out} \mid)$ MeV, as it is more
realistic~\cite{MKS00}.
\par

\subsection{Role of Different Skyrme Forces}

Figs.~2(a) and (b) shows the fragmentation potential $V(\eta)$ and
fractional yield at $R = R_{min}$ with $V(R_{min})=
\overline{V}_{min}$. The classical hydrodynamical mass parameter
$B_{\eta \eta}$ of Kr\"oger and Scheid~\cite{kro80} used in the
calculation of preformation probability. The fractional yields are
calculated within PCM at $T$ = 3.0 MeV using different Skyrme
forces for $^{56}$Ni$^{\ast}$. From the figure, we observe that
different Skyrme forces do not alter the transfer structure of
fractional yields. The Skyrme force parameters have marginal role
to play. Some variations in the absolute values are however
visible~\cite{NkConf}. The fine structure is not at all disturbed
for different sets of Skyrme forces.
\par

The results for the cluster decay half-lives in $^{56}$Ni$^{\ast}$
are quantified by the following quantity as:
\begin{equation}
\left[\log T_{1/2} \right] \% = \frac{(\log T_{1/2})^i- (\log
T_{1/2})^{SIII}}{(\log T_{1/2})^{SIII}}\times 100, \label{eq:34}
\end{equation}
where $i$ stands for different sets of Skyrme force parameters and
SIII for one set of Skyrme force parameters, which is widely used.
Here, the strength parameter of surface correction is taken as
zero (i.e. $\lambda =0$).
\par

In Fig.~3(a) and (b), we display the quantified results using
Eq.~(\ref{eq:34}) for $\log  T_{1/2}$ within PCM and UFM models as
a function of cluster mass $A_2$. The role of temperature $T$  (or
excitation energy $E_{CN}^{\ast}$) enters only in the PCM via
preformation probability $P_0$. These variation in the cluster
decay half-lives for different clusters lies within $\pm$10\% for
PCM and $\pm$15\% for UFM. This amount is significant once we
understand cluster decay probabilities can be measured with great
accuracy in the literature.

\subsection{Role of Strength Parameter of Surface Correction
($\lambda$)}

The effect of different $\lambda$-values for the heavy-ion nuclear
potential is analyzed in Refs.~\cite{8Puri92,8Puri06}, suggesting
that different $\lambda$-value, can alter the depth of the nuclear
potential $V_N$ significantly.  In Ref.~\cite{8Puri92}, it was
shown that the barrier heights gets lowered whereas the fusion
barrier position  shifts outward where stronger role of $\lambda$
is taken into account. The effect of this strength parameter
$\lambda$ for additional surface effects in the decay calculations
has yet not been studied in the literature. In this subsection, we
plan to study the effect of strength parameter of surface
correction on cluster decay half-lives by taking different
$\lambda$-values (equal to $0,~1/36,~2/36,~3/36,~4/36$, and
$5/36$) in SEDM for the compound system $^{56}$Ni$^{\ast}$.
\par

In Fig. 4, the scattering potential for different values of
surface correction factor $\lambda$ is plotted as a function of
internuclear distance $R$. One observes from the figure that
variation in the $\lambda$-value changes the interior part of the
scattering potential thereby changing the penetration probability.
\par

In Fig 5(a) and (b), we show the fragmentation potential
$V(\eta)$and fractional mass distribution yield at
 $R = R_{min}$ with $V(R_{min})= \overline{V}_{min}$. The fractional yields are calculated within PCM
at $T$ = 3.0 MeV using different values of surface correction
factor for $^{56}$Ni$^{\ast}$. From figure, we observe that
different values of $\lambda$ changes the fractional yield to
large extent but do not alter its transfer structure. The fine
structure is not at all disturbed for different values of surface
correction factor.
\par

The results for the cluster decay half-lives in $^{56}$Ni$^{\ast}$
are quantified by the following quantity as:
\begin{equation}
\left[\log T_{1/2} \right] \% = \frac{(\log T_{1/2})^i- (\log
T_{1/2})^{\lambda=0}}{(\log T_{1/2})^{\lambda=0}}\times 100,
\label{eq:35}
\end{equation}
where $i$ stands for different $\lambda$-values of the strength
parameter of surface correction. Skyrme force SIII is employed for
these calculations. In Fig.~6, we display the quantified results
using Eq.~(35) for the percentage variation of $\log  T_{1/2}$
within PCM and UFM as a function of cluster mass $A_2$. The
variation in the cluster decay half-lives for different clusters
lies within $\pm$10\% for both PCM and UFM. Together with the
effect of different Skyrme forces, one can see that the net effect
of different Skyrme forces as well as surface corrections has
sizable effect on the cluster decay half-life times.
\section{\label{summary}Summary}
We here  reported the role of different Skyrme forces as well as
surface corrections in the cluster decay constant calculations.
Our studies revealed that the effect of different Skyrme forces on
the cluster decay half-life times is about $\pm$15\%, whereas it
is $\pm$10\% in the case of surface corrections. \\

 This work was supported by a research grant from the Department of
Atomic Energy, Government of India.


\newpage
\begin{table}[h]
  \centering
  \caption{The calculated characteristic quantities for cluster
decay of $^{56}$Ni$^{\ast}$ compound system  for fragment masses
$A_2 \geq 16$, with excitation energies  $E^{\ast} = Q_{eff} +
\mid Q_{out} \mid$.  }
\vskip 1.5cm
\begin{tabular}{ l  c  c  c    }\hline
Cluster +  & $\mid Q_{out} \mid$ & $Q_{eff}$& $E^{\ast}$   \\
   Daughter & (MeV)  &  (MeV) &(MeV)   \\ \hline
$^{16}$O + $^{40}$Ca &-14.12   &5.55 &19.67   \\\hline

$^{18}$Ne + $^{38}$Ar &-22.23    &2.31 &24.54   \\\hline

$^{20}$Ne + $^{36}$Ar &-17.12    &4.35 &21.47   \\\hline

$^{22}$Mg + $^{34}$S &-24.58    &1.37 &25.95   \\\hline

$^{24}$Mg + $^{32}$S &-16.57    &4.57 &21.14   \\\hline

$^{26}$Si + $^{30}$Si &-23.57    &1.77 &25.34   \\\hline

$^{28}$Si + $^{28}$Si &-12.20    &6.32 &18.52   \\\hline
\end{tabular}
\end{table}

\newpage

\begin{figure}
\centering
\includegraphics* [scale=0.7]{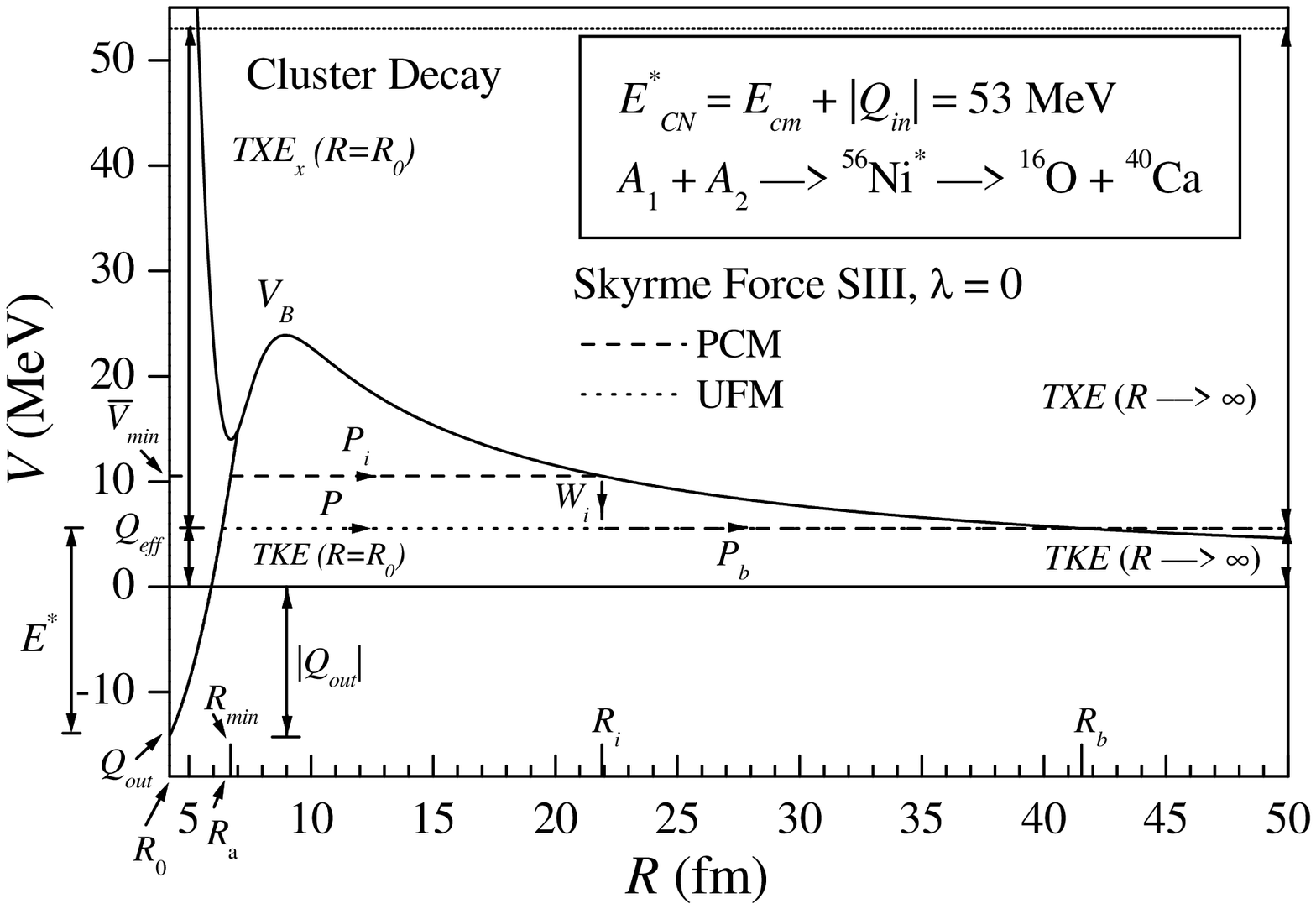}
\vskip -10.0 cm \caption {The scattering potential $V(R)$ (in MeV)
for cluster
  decay of $^{56}$Ni$^{\ast}$ into $^{16}$O + $^{40}$Ca channel using Skyrme force SIII, with $\lambda=0$. The
  distribution of compound nucleus excitation energy E$_{CN}^{*}$
  at both the initial ($R=R_{0}$) and asymptotic ($R \to  \infty$)
  stages and $Q$-values are shown. The decay path for both PCM and
  UFM models is also displayed.}
\end{figure}
\begin{figure}
\centering
\includegraphics* [scale=0.7]{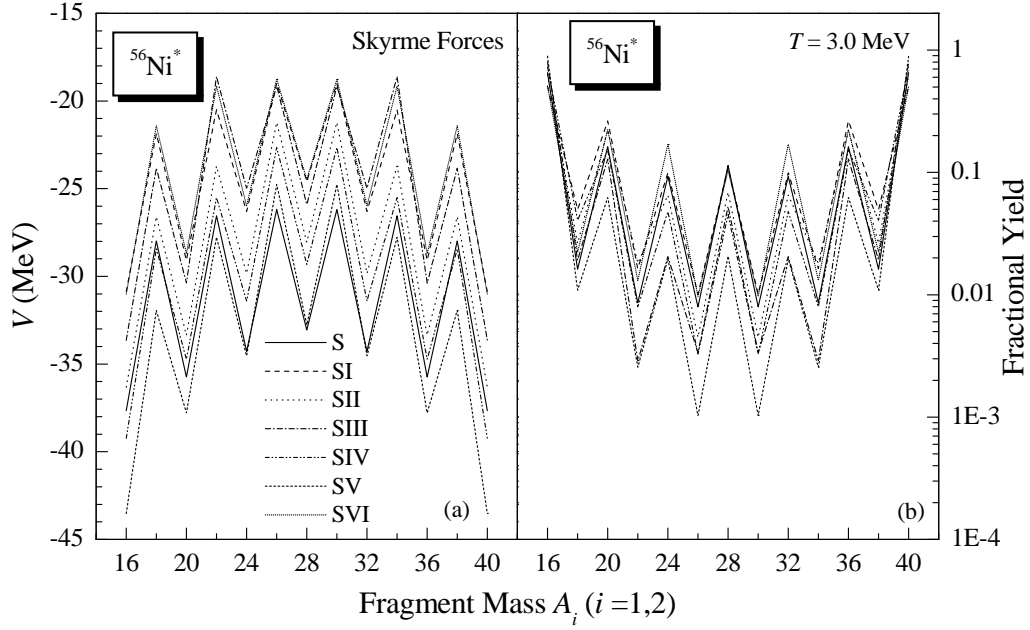}
\vskip -10.0 cm
 \caption {(a) The fragmentation potential $V(\eta)$
and (b) calculated fission mass distribution yield with different
Skyrme forces  at $T$ = 3.0 MeV.}
 \vskip -0.4 cm
\end{figure}
\begin{figure}
\centering
\includegraphics* [scale=0.7]{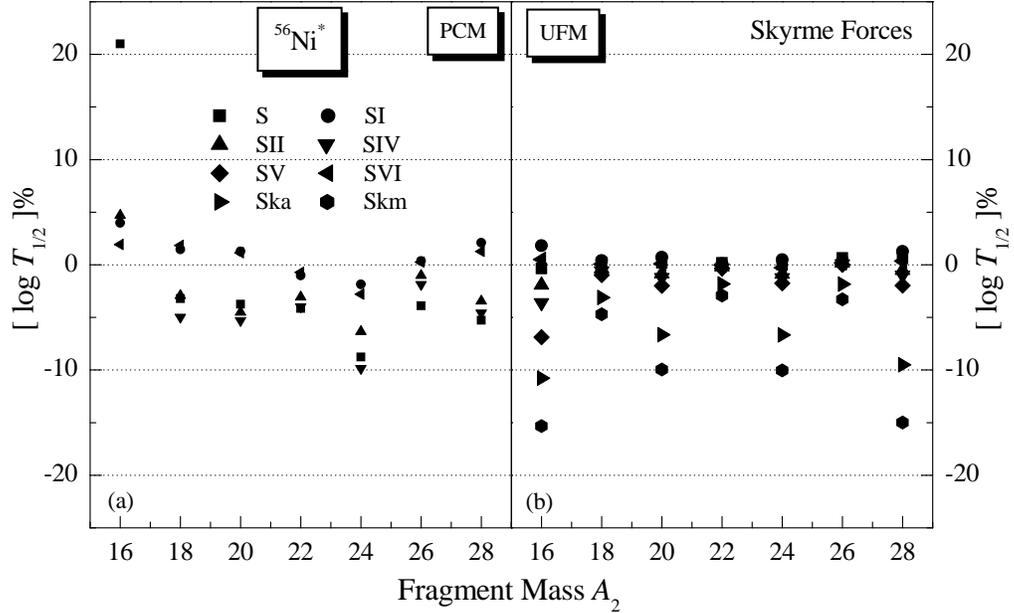}
\vskip -10.0 cm
\caption {Percentage variation of $\log  T_{1/2}$
for different Skyrme forces  w.r.t. SIII force.} \vskip -0.4 cm
\end{figure}
\begin{figure}
\centering
\includegraphics* [scale=0.6]{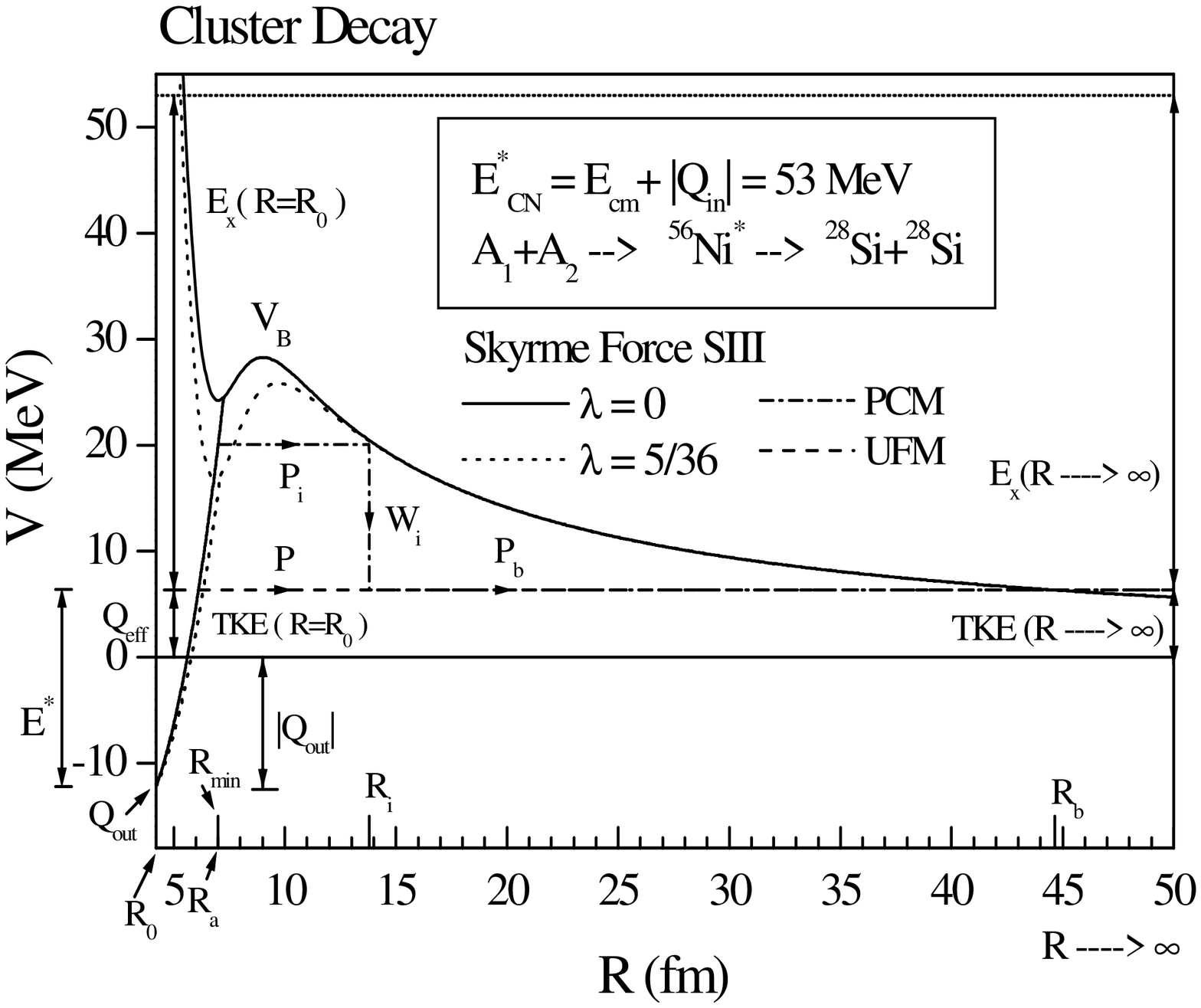}
\vskip -7.0 cm \caption {Same as Fig~1, but for different  values
of surface correction factor ($\lambda$).} \vskip -0.4 cm
\end{figure}
\begin{figure}
\centering
\includegraphics* [scale=0.7]{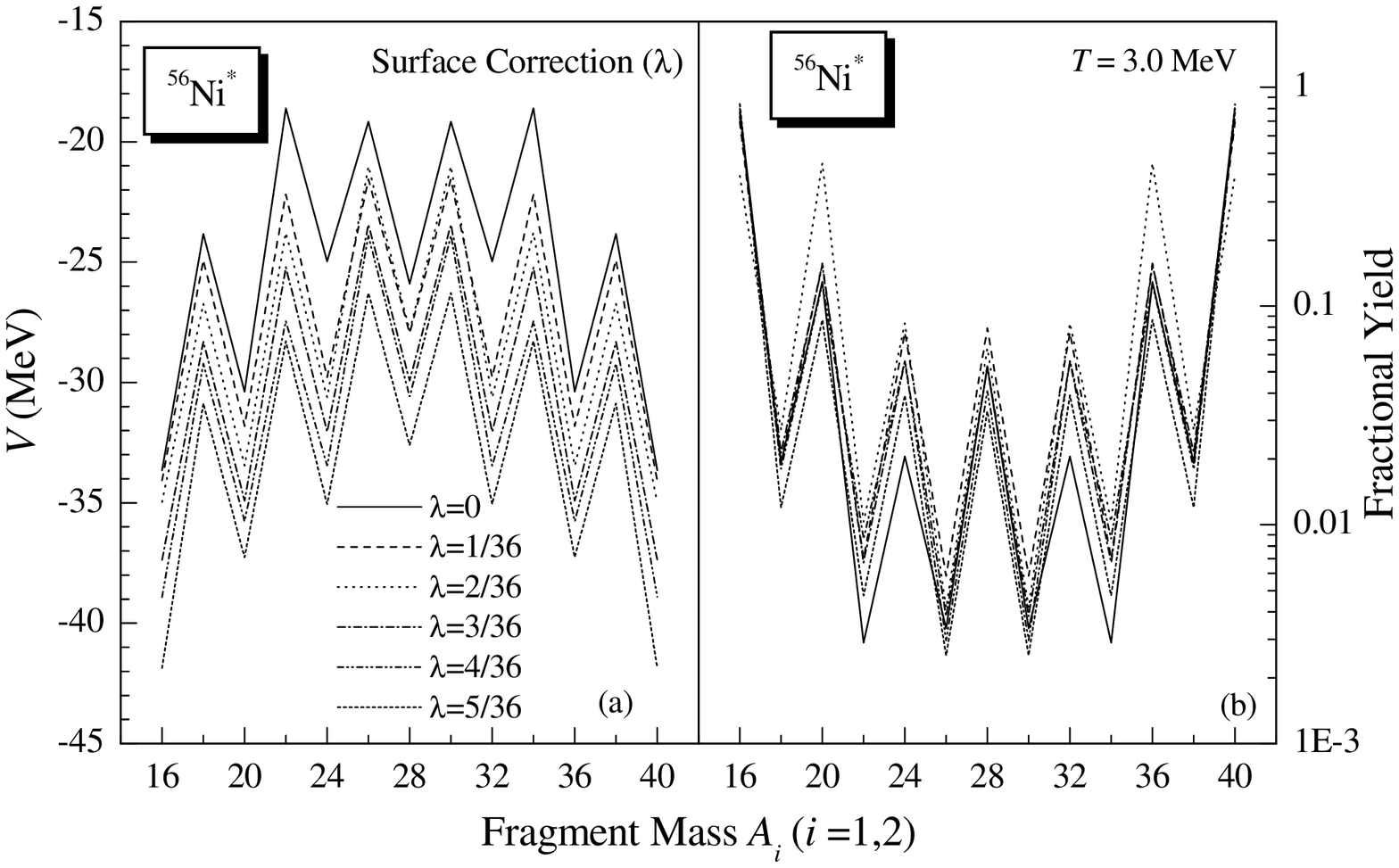}
\vskip -10.0 cm
\caption {Same as Fig~2, but for different values
of surface correction factor ($\lambda$).} \vskip -0.4 cm
\end{figure}
\begin{figure}
\centering
\includegraphics* [scale=0.7]{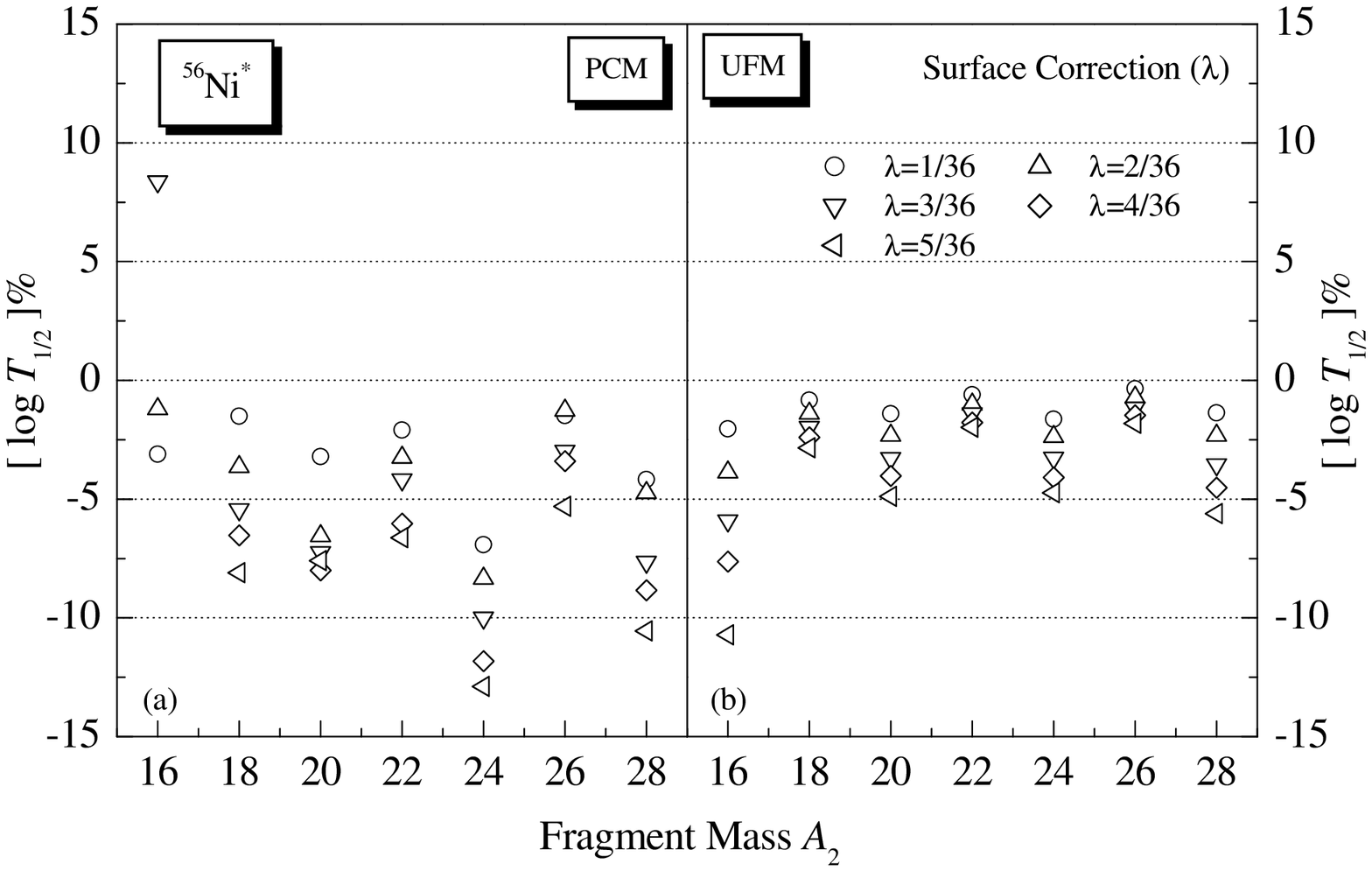}
\vskip -10.0 cm
\caption {Same as Fig~3, but for different values
of strength parameter of surface correction ($\lambda \neq 0$)
w.r.t. surface correction ($\lambda = 0$).}
 \vskip -0.4 cm
\end{figure}


\begin{thebibliography}{0}
\bibitem{id} J.M.B. Shorto \emph{et al.}, {\it Phys. Rev.}
\textbf{C81,} 044601 (2010); I. Dutt, R.K. Puri {\it ibid.}
\textbf{81,} 047601 (2010); {\it ibid.} \textbf{81,} 044615
(2010); {\it ibid.} \textbf{81,} 064609 (2010); {\it ibid.}
\textbf{81,} 064608 (2010).

\bibitem{qmd}C. Xu, B.A. Li {\it Phys. Rev.}
\textbf{C 81,} 044603 (2010); S. Kumar {\it ibid.} \textbf{78,}
064602 (2008); {\it ibid.} \textbf{81,} 014611 (2010); {\it ibid.}
\textbf{81,} 014601 (2010); Y.K. Vermani  \emph{et al.}, {\it J.
Phys. G: Nucl. Part. Phys.} \textbf{36,} 105103 (2010); {\it
ibid.} \textbf{37,} 015105 (2010); {\it ibid.} {\it Europhys.
lett.} \textbf{85,} 62001 (2009); {\it ibid.} {\it Phys. Rev.}
\textbf{C 79,} 064613 (2009); A. Sood {\it et al.}, {\it ibid.}
\textbf{79,} 064618 (2009); S. Gautam {\it et al.}, {\it J. Phys.
G: Nucl. Part. Phys.} \textbf{37,} 085102 (2010); Y. Vermani, R.K.
Puri {\it Nucl. Phys.} \textbf{A,} (2010) in press.


\bibitem{gupta} S.K. Patra \emph{et al.}, {\it Phys. Rev.}
\textbf{C 80,} 034612 (2009); S.K. Arun \emph{et al.}, {\it ibid.}
\textbf{80,} 034317 (2009); {\it ibid.} \textbf{79,} 064616
(2009); R. Kumar \emph{et al.}, {\it ibid.} \textbf{79,} 034602
(2009).

\bibitem{kp} K.P. Santhosh \emph{et al.}, {\it J. Phys. G: Nucl. Part. Phys.} \textbf{36,}
 115101 (2009); \emph{ibid.} \textbf{36,}
 015107 (2009);  {\it Pramana J. Phys.} {\bf 59,} 599 (2002).
\bibitem{7Poen} D.N. Poenaru, W. Greiner, R. Gherghescu, {\it Phys. Rev.} \textbf{C47,} 2030
(1993); H.F. Zhang \emph{et al.}, {\it ibid.} \textbf{80,} 037307
(2009).
\bibitem{7Buck} R.K. Gupta \emph{et al.}, {\it J. Phys. G: Nucl. Part. Phys.} \textbf{26,} L23 (2000); B. Buck, A.C. Merchant, S.M. Perez, {\it Nucl. Phys.} \textbf{A512,} 483 (1990);
B. Buck, A.C. Merchant, {\it J. Phys. G: Nucl. Part. Phys.}
\textbf{16,} L85 (1990).

\bibitem{7Sand} A. Sandulescu \emph{et al.}, {\it Int. J. Mod. Phys.} \textbf{E1,} 379 (1992); R.K. Gupta, \emph{et al.}, {\it J. Phys. G: Nucl. Part. Phys.} \textbf{19,} 2063
(1993); {\it Phys. Rev.} \textbf{C56,} 3242 (1997).
\bibitem{rkg88} R.K. Gupta, {\it 5th International Conference on Nuclear Reaction Mechanisms, Varenna, Italy}, p. 416 (1988).
\bibitem{mal89} S.S. Malik, R. K. Gupta, {\it Phys. Rev.} \textbf{C39,} 1992
(1989); \emph{ibid.} \textbf{C50,} 2973 (1994); S.S. Malik  {\it
et al.},  {\it Pramana J. Phys.} {\bf 32,} 419 (1989); R.K. Puri
{\it et al.}, {\it Europhys. Lett.} {\bf 9,} 767 (1989); R. K.
Puri \emph{et al.}, {\it J. Phys. G: Nucl. Part. Phys.}
\textbf{18,} 903 (1992).
\bibitem{kum97} S. Kumar, R.K. Gupta, {\it Phys. Rev.} \textbf{C55,} 218 (1997).
\bibitem{8MS} W.D. Myers, W.J. \'Swiatecki, {\it Phys. Rev.} \textbf{C62,}  044610 (2000).
\bibitem{8CW} P.R. Christensen, A. Winther, {\it Phys. Lett.} \textbf{B65,} 19 (1976).
\bibitem{8Blocki77} J. Blocki \emph{et al.}, {\it Ann. Phys.}  \textbf{105,} 427 (1977).
\bibitem{8Bass} R. Bass, {\it Nucl. Phys.} \textbf{A231,} 45 (1974); {\it Phys. Rev. Lett.} \textbf{39,} 265  (1977).
\bibitem{8Deni02} V.Y. Denisov, {\it Phys. Lett.} \textbf{B526,} 315 (2002); V.Y. Denisov, S. Hofmann, {\it Phys. Rev.} \textbf{C61,} 034606 (2000); \emph{ibid.} \textbf{76,} 014602 (2007); \emph{ibid.}
\textbf{81,} 034613 (2010); \emph{ibid.} \textbf{81,} 025805
(2010).
\bibitem{8Stancu} D.M. Brink, F. Stancu, {\it  Nucl. Phys.} \textbf{A243,} 175 (1975); F. Stancu, D.M. Brink, \emph{ ibid}, \textbf{A270,} 236 (1976).
\bibitem{8Puri92} R.K. Puri \emph{et al.}, {\it Eur. Phys. J.} \textbf{A23,}
429 (2005); R. Arora \emph {et al.}, {\it ibid.} \textbf{8,} 103
(2000); R.K. Puri \emph {et al.}, {\it ibid.} \textbf{3,} 277
(1998); R.K. Puri \emph {et al.}, {\it Phys. Rev.} \textbf{C51,}
1568 (1995); {\it ibid.} \textbf{45,} 1837 (1992);  {\it ibid.}
\textbf{43,} 315 (1991); {\it ibid.} {\it J. Phys. G: Nucl. Part.
 Phys.} \textbf{18,} 903 (1992); R.K. Puri, R.K. Gupta, {\it Int. J. Mod. Phys.} \textbf{E1,} 269 (1992).
\bibitem{8Dob} A. Dobrowolski, K. Pomorski, J. Bartel, {\it  Nucl. Phys.} \textbf{A723,} 93
(2003); J. Bartel  {\it et al.}, {\it Eur. Phys. J.} {\bf A14,}
179 (2002); M. Liu \emph {et al.}, {\it Nucl. Phys.}
\textbf{A768,} 80 (2006); N. Wang, J.Q. Li, E.G. Zhao, {\it Phys.
Rev.} {\bf C74,} 044604 (2006).
\bibitem{8Brack} M. Brack, C. Guet, H.-B. H{\aa}kansson, {\it Phys. Rep.} \textbf{123,} 275 (1985).
\bibitem{2Skyrme} T.H.R. Skyrme, {\it  Phil. Mag.} \textbf{1}, 1043 (1956); {\it Nucl. Phys.} \textbf{9,}  615 (1959).
\bibitem{VB72} D. Vautherin, D.M. Brink, {\it Phys. Rev.} \textbf{C5,} 626 (1972).
\bibitem{shen09} Q. Shen, Y. Han, H. Guo, {\it Phys. Rev.} \textbf{C80,} 024604
 (2009); Z. Feng, G. Jin, F. Zhang, {\it Nucl. Phys.} \textbf{A802,} 91
 (2008).
\bibitem{sk} S. Kumar \emph{et al.},  {\it Phys. Rev.} \textbf{C58,} 3494 (1998); {\it ibid.}
\textbf{C58,} 1618 (1998); J. Singh  \emph{et al.}, {\it Phys.
Rev.} \textbf{C62,} 044617 (2000); J. Dhawan \emph{et al.}, {\it
Phys. Rev.} \textbf{C75,} 057601 (2007); R.K. Puri \emph{et al.},
{\it Nucl. Phys.} \textbf{A575,} 733 (1994); D.T. Khoa \emph{et
al.}, {\it Nucl. Phys.} \textbf{A542,} 671 (1992); S.W. Huang
\emph{et al.}, {\it Phys. Lett.} \textbf{B298,} 41 (1993); G.
Batko \emph{et al.}, {\it J. Phys. G: Nucl. Part. Phys.}
\textbf{20,} 461 (1994); S.W. Huang et al., {\it Prog. Part. Nucl.
Phys.} \textbf{30,} 105 (1993); E. Lehmann et al., {\it Prog.
Part. Nucl. Phys.}  \textbf{30,} 219 (1993).
\bibitem{rk}  R.K. Puri \emph{et al.},  {\it Phys. Rev.} \textbf{C54,} R28 (1996); {\it ibid.}  {\it J. Comput. Phys.} \textbf{162,} 245 (2000);
A. Sood et al., {\it Phys. Rev.} \textbf{C70,} 034611 (2004); {\it
ibid. Phys. Lett.} \textbf{B594,} 260 (2004); P.B. Gossiaux {\it
Nucl. Phys.} \textbf{A619,} 379 (1997); C. Fuchs et al., {\it J.
Phys. G: Nucl. Part. Phys.} \textbf{22,} 131 (1996); E. Lehmann
\emph{et al.}, \emph{Z. Phys} \textbf{A355} 55 (1996); E. Lehmann
\emph{et al.}, {\it Phys. Rev.} \textbf{C51,} 2113 (1995); S.
Kumar \emph{et al.},  {\it Phys. Rev.} \textbf{C58,} 320 (1998);
{\it Phys. Rev.} \textbf{C57,} 2744 (1998).

\bibitem{Li91} G. Lingxiao, Z. Yizhong, W. N\"orenberg, {\it Nucl. Phys.} \textbf{A459,} 77 (1986); Li G. Qiang, {\it J. Phys. G: Nucl. Part. Phys.} \textbf{17,} 1 (1991).
\bibitem{Sanders99} S.J. Sanders, A. S. de Toledo, C. Beck, {\it Phys. Rep.} \textbf{311,} 487 (1999).
\bibitem{Sanders88} S.J. Sanders \emph{et al.}, {\it Phys. Rev. Lett.} \textbf{59,} 2856 (1987); {\it ibid.} \textbf{61,} 2154 (1988).
\bibitem{Sanders91} S.J. Sanders, {\it Phys. Rev.} \textbf{C44,} 2676 (1991).
\bibitem{Nouicer99} R. Nouicer \emph{et al.}, {\it Phys. Rev.} \textbf{C60,} 041303 (1999).
\bibitem{Beck01} C. Beck \emph{et al.}, {\it Phys. Rev.} \textbf{C63,} 014607 (2001).
\bibitem{Thumm01} S. Thummerer \emph{et al.}, {\it J. Phys. G: Nucl. Part. Phys.} \textbf{27,} 1405 (2001).
\bibitem{Bhatt02} C. Bhattacharya \emph{et al.}, {\it Phys. Rev.} \textbf{C57,} 203 (2001); {\it ibid.} \textbf{C65,} 014611 (2002).
\bibitem{Sanders86} S.J. Sanders \emph{et al.}, {\it Phys. Rev.} \textbf{C34,} 1746 (1986).

\bibitem{Betts} R.R. Betts, {\it Conference on Resonances in Heavy Ion Reactions, Bad H\"onnef: Lecture Notes in Physics}, Vol. \textbf{156,} edited by K. A. Eberhardt, Springer, Berlin, p. 185 (1981).
\bibitem{Sanders89} S.J. Sanders \emph{et al.}, {\it Phys. Rev.} \textbf{C40,} 2091 (1989).
\bibitem{Sanders94} S.J. Sanders \emph{et al.}, {\it Phys. Rev.} \textbf{C49,} 1016 (1994).
\bibitem{nkd03} R.K. Gupta \emph{et al.}, {\it Phys. Rev.} \textbf{C68,} 014610 (2003).
\bibitem{MKS00} M.K. Sharma, R.K. Gupta, W. Scheid, {\it J. Phys. G: Nucl. Part. Phys.} \textbf{26,} L45 (2000).
\bibitem{mat97} T. Matsuse \emph{et al.}, {\it Phys. Rev.} \textbf{C55,} 1380 (1997).
\bibitem{shiva87} B. Shivakumar \emph{et al.}, {\it Phys. Rev.} \textbf{C35,} 1730 (1987).

\bibitem{gupta85} P. Chattopadhyay, R.K. Gupta, {\it Phys. Rev.} \textbf{C30,} 1191
 (1984), and earlier references therein.
\bibitem{2Von} C.F. von Weizs\"acker, {\it Z. Phys.} \textbf{96,} 431 (1935).
\bibitem{Elton} L.R.B. Elton, \emph{Nuclear sizes}, {\it Oxford University Press, London} (1961); H.de Vries, C.W. de Jager, C.de Vries, {\it At. Data Nucl. Data Tables} \textbf{36,} 495 (1987).
\bibitem{7Puri} R.K. Gupta \emph{et al.}, {\it J. Phys. G: Nucl. Part. Phys.} \textbf{26,} L23 (2000).
\bibitem{rkg75}R.K. Gupta, W. Scheid, W. Greiner, {\it Phys. Rev. Lett.} \textbf{35,} 353 (1975).
\bibitem{7SNG} D.R. Saroha, N. Malhotra, R.K. Gupta, {\it J. Phys. G: Nucl. Part. Phys.} \textbf{11,} L27 (1985).
\bibitem{7Maruhn74} J. Maruhn, W. Greiner, {\it Phys. Rev. Lett.} \textbf{32,} 548 (1974).
\bibitem{mol95}P. M\"oller \emph{et al.}, {\it At. Data Nucl. Data Tables} \textbf{59,} 185 (1995).
\bibitem{kro80} H. Kr\"oger, W. Scheid, {\it J. Phys. G: Nucl. Part. Phys.} \textbf{6,} L85 (1980).
\bibitem{7Rkg94} R.K. Gupta, W. Greiner, {\it Int. J. Mod. Phys.} \textbf{E3,} 335 (1994).
\bibitem{NkConf} N.K. Dhiman, R.K. Puri, {\it 1st Chandigarh Science Congress}, Vol. \textbf{IA,} p. 226, March 10-11 (2007).
\bibitem{8Puri06} N.K. Dhiman, R.K. Puri, {\it Acta Phys. Pol.} \textbf{B37,} 1855 (2006).
\end{thebibliography}
\end{document}